\documentclass[runningheads]{llncs}
\usepackage{graphicx}

\def\BibTeX{{\rm B\kern-.05em{\sc i\kern-.025em b}\kern-.08em
    T\kern-.1667em\lower.7ex\hbox{E}\kern-.125emX}}

\usepackage{cite}
\usepackage[hyphens,spaces,obeyspaces]{url}
\usepackage{amsmath,amssymb,amsfonts}
\usepackage{algorithmic}
\usepackage{textcomp}
\usepackage{xcolor}
\usepackage{hyperref}

\newif\ifediting
\editingtrue  %

\newif\iflong %

\definecolor{mygray}{gray}{0.2}
\definecolor{mycyan}{rgb}{0.05, 0.5, 0.5}

\ifediting
    \usepackage{comment}
    \usepackage[draft,textsize=footnotesize,textwidth=17mm]{todonotes}
\fi

\def\inlinedremark#1#2#3{        
  \ifediting
        \textcolor{#2}{ #3}
    \else 
        {#3}
    \fi
    }

\def\CScom#1{\remark{Cristiana}{cyan!30}{#1}}
\def\NBcom#1{\remark{Nataliia}{orange!30}{#1}}

\def\MTcom#1{\remark{Michael}{gray!30}{#1}}

\def\CStext#1{\inlinedremark{Cristiana}{violet!60}{#1}}
\def\NBtext#1{\inlinedremark{Nataliia}{orange!100}{#1}}

\def\REMARK#1#2{        %
  \ifediting
    \begin{center}
    \noindent\fbox{
        \begin{minipage}{0.9\textwidth}
            \textrm{{[#1]: #2}}
        \end{minipage}}
    \end{center}
    \fi
    }

\def\NBcom#1{\textbf{\REMARK{Nataliia}{\textcolor{orange!100}{#1}}}}
\def\MTcom#1{\textbf{\REMARK{Michael}{\textcolor{cyan!90}{#1}}}}

\def\CScom#1{\textbf{\REMARK{Cristiana}{\textcolor{violet!60}{#1}}}}

\newif\iffull %
\fulltrue   %

\begin{document}%
\title{Consent Management Platforms under the GDPR: processors and/or controllers?\thanks{A preliminary version of this paper is presented for discussion only, with no official proceedings at ConPro'21: \url{https://www.ieee-security.org/TC/SPW2021/ConPro/}.}
}
\author{
Cristiana Santos\inst{2}
\and
Midas Nouwens\inst{3}
\and 
Michael Toth\inst{1}
\and
Nataliia Bielova\inst{1}
\and
Vincent Roca\inst{1}
}
\authorrunning{C. Santos et al.}
\institute{Inria, France\\
\email{\{nataliia.bielova,michael.toth,vincent.roca\}@inria.fr}
\and
Utrecht University, The Netherlands\\
\email{c.teixeirasantos@uu.nl}
\and
Aarhus University, Denmark\\
\email{midasnouwens@cc.au.dk}}

\maketitle              %

\def\consentbanner{consent pop-up}
\def\consentsignal{Consent Signal}

\begin{abstract}

Consent Management Providers (CMPs) provide \consentbanner s that are embedded in ever more websites over time to enable streamlined compliance with the legal requirements for consent mandated by the ePrivacy Directive and the General Data Protection Regulation (GDPR). They implement the standard for consent collection  from the Transparency and Consent Framework (TCF) (current version v2.0) proposed by the European branch of the Interactive Advertising  Bureau (IAB Europe). Although the IAB’s TCF specifications  characterize CMPs as data processors, CMPs factual activities often qualifies them as data controllers instead. Discerning  their clear role is crucial since compliance obligations and CMPs liability depend on their accurate characterization. We perform empirical  experiments  with two major CMP providers in the EU: Quantcast and OneTrust and paired with a legal analysis. We conclude that CMPs  process  personal  data, and we identify multiple scenarios wherein CMPs are controllers. 

%

%

%

%

%

%

\iffalse

\ifediting
To remove all comments at once, please comment out the line "$\backslash$editingtrue".

Example comment that will disappear in submission (use your name for your comment):
\CScom{comment}
%
%
%
%
Example for new inlined text that will remain in submission without color and wihtout name:\CStext{text} 
%
%
\fi

\fi

\end{abstract}

\section{Introduction}
To comply with the General Data Protection Regulation (GDPR)~\cite{GDPR} and the ePrivacy Directive (ePD)~\cite{ePD-09}, a website owner 
needs to first obtain \emph{consent} from users, and only then is allowed to %
process personal data when offering goods and services and/or monitoring the users’ behavior. 
\def\freestyle{free-style}
As a result, %
numerous companies have started providing ``\emph{Consent as a Service}'' solutions to help website owners ensure legal compliance ~\cite{Sant-etal-20-TechReg}. 

\begin{figure}[!tpb]
    \centering
    \includegraphics[width=0.95\textwidth]{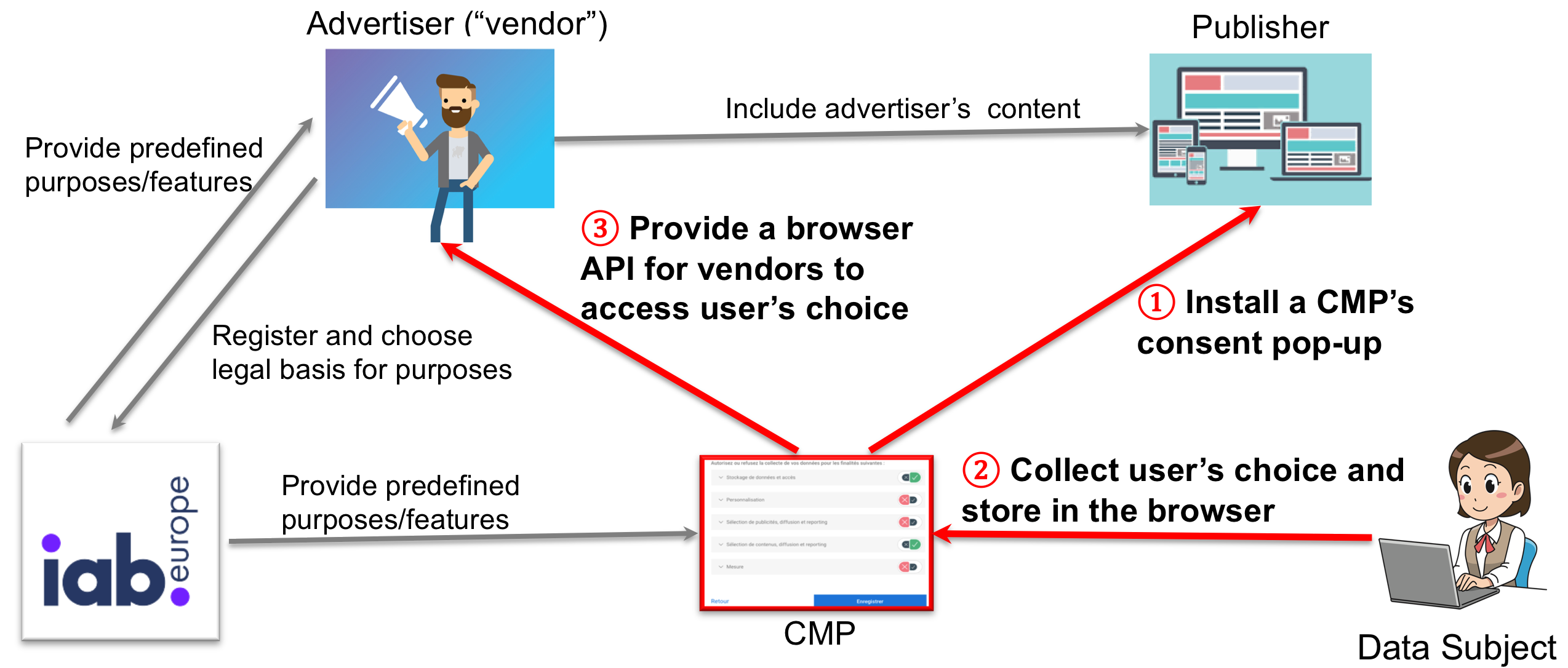}
    \caption{\textbf{Actors under IAB Europe TCF ecosystem:} IAB Europe, Advertisers (called “vendors”), Consent Management Providers (CMPs), Publishers, Data Subjects. The IAB Europe defines the purposes
and features that are shown to users. Registered vendors declare purposes and legal basis and the features upon which they rely. CMPs provide consent pop-up, store the user’s choice as a browser cookie,
and provide an API for advertisers to access this information.}
    \label{fig:CMPs}
\end{figure}

To standardise\footnote{Standardization is used within the meaning of streamline at scale consent implementation.} the technical implementation of these consent pop-ups, the European branch of the Interactive Advertising Bureau (IAB Europe), an industry organisation made up of most major advertising companies in the EU, %
developed a Transparency and Consent Framework (TCF)~\cite{IAB-TCF2020}. 
This framework (currently on version 2.0) was developed to preserve the exchange of data within the advertising ecosystem, which now requires being able to demonstrate how, when, from who, and on which legal basis that data is collected.
The actors in this ecosystem are IAB Europe, advertisers (called ``vendors''), Consent Management Providers (CMPs), publishers, and data subjects (see Figure~\ref{fig:CMPs}).
%
%
%
%
%
%

%
%
%
%
%

%
%
%
%

%
%
%
\iffalse 
IAB Europe TCF is a technical standard ``to help all parties in the digital advertising chain ensure that they comply with the EU’s GDPR and ePrivacy Directive when processing personal data or accessing and/or storing information on a user’s device''~\cite{IAB-TCF-website} 
\fi

%

%

%

%
%
%
%
%
%
%
%
%

%
%
%
%
%
%
%

%
%
\iffalse
The following %
observations motivate us to %
focus our study on the role of CMPs.
 First, CMPs are embedded in more and more websites over time~\cite[Fig. 6]{Hils-etal-20-IMC}. Currently there are %
115 unique CMPs %
in the IAB Europe TCF.~\footnote{This observation is based on a CMP list accessed on January 14, 2021: \url{https://iabeurope.eu/cmp-list/}.}
Second, CMPs are responsible for the choices of consent design configurations that they offer to publishers -- CMPs have the capacity to control data collection and to standardize the options afforded to users. Through design of consent pop-ups, 
CMPs can even restrict access to a website through design practices like 
consent wall and tracking walls~\cite{Gray-etal-21-CHI-2}. 
Finally, CMPs are in unique position to include any other tools (such as tracking techniques) in the consent mechanisms and affect at once all websites that include such CMP and all users that visit the website. 
\fi

%
%
%
%
%
\iffalse
\emph{Consent Management Providers (CMPs)} are third-party services which collect consent from website visitors through interactive interfaces and redistribute it to advertisers.
%
%
The following %
observations motivate us to %
focus our study on the role of CMPs.
\fi

Although recent work has started to address the complex technical and legal aspects of the IAB Europe TCF ecosystem~\cite{Hils-etal-20-IMC,Matt-etal-20-IEEESP, Matt-etal-20-APF, Nouwens-etal-20-CHI, Pawlata2020, Dege-etal-19-NDSS,feedback-IAB-EDPB}, \emph{neither prior work nor court decisions} have so far discussed the role of the CMPs.
Therefore,  it is currently unclear what the role of these CMPs is under the GDPR,
and consequently what their legal requirements and liabilities are.

\iffull
\else
If a CMP is established as a data processor and fails to comply with its obligations under the GDPR, then it can be held liable and fined (Articles 28(3)(f) and 32-36 GDPR). Moreover, if a false
\consentsignal\ 
is stored and transmitted, it may well be 
considered an
``unauthorised disclosure of, or access to personal data transmitted, stored or otherwise processed''
~\cite[Art. 32(2)]{GDPR}. If instead a CMP %
is a controller, it is required to obtain personal data fairly, lawfully, and in compliance with any transparency requirements with respect to users. A breach of these obligations will make a CMP liable to sanctions (Article 28(10)). 
\fi 

This paper examines if and when CMPs can be considered a \textit{data controller} -- i.e., an actor responsible for determining the purposes and means of the processing of personal data (Art. 4(7) GDPR) -- or a \textit{data processor} -- i.e., an actor which processes personal data on behalf of the controller (Art. 4(8) GDPR).

Discerning the correct positioning of CMPs is crucial since compliance measures and CMPs liability depend on their accurate characterization (GDPR Recital 79).
%
%
%
%
%
%
%
%
%
%
%
%
%
%
\iffalse 
i) \emph{Prevalence}. CMPs are embedded in ever more websites over time~\cite[Fig. 6]{Hils-etal-20-IMC}. Currently there are %
115 unique CMPs %
in the IAB Europe TCF.~\footnote{This observation is based on a CMP list accessed on January 14, 2021: \url{https://iabeurope.eu/cmp-list/}.}
 \NBcom{See a list of all CMPs here on overleaf: cmp-sources folder: 2021-01-14-cmp-list.xlsx}
 
ii) \emph{Business model}. The consent ecosystem disposed by CMPs has been established as a business model based on personal data between websites and third-party vendors~\cite{Hils-etal-20-IMC}; 

iii) \emph{Standardizing design interfaces of consent pop-ups}.
CMPs propose design configurations to publishers:
%
they have the capacity to control the data collection and to standardize the options afforded to users (a de facto standard)~\cite{Hils-etal-20-IMC}.
CMPs can even restrict access to a website through design practices like %
consent wall and tracking walls~\cite{Gray-etal-21-CHI-2}. 
%

%

iv) \emph{Accountability}. Attributing the role of controller is to ensure accountability and the
effective and comprehensive protection of the personal data~\cite{EDPB-controller-2020}. %
\fi
%
To determine the role of CMPs under the GDPR, in this paper we answer the following research questions:
\begin{description}
    \item[\S\ref{sec:CMPs}] When  are  CMPs  processing  personal data? %
    \item [\S\ref{sec:processors}] When  do CMPs act as data  processors? %
    \item [\S\ref{sec:controllers}] When do CMPs act as data controllers? %
\end{description}

Note that the TCF is a voluntary framework: not all CMPs are part of it and abide by its policies. 
However, it has become a \textit{de facto} standard used by a growing number of actors ~\cite[Fig. 6]{Hils-etal-20-IMC}.
This means that focusing on the CMPs within this ecosystem provides results that can more easily be generalised, compared to looking at the specific %
implementations of individual CMPs.
Whenever we refer to CMPs in the rest of the article, we are referring to CMPs registered as part of the IAB Europe TCF.
Our argumentation is based on: 
\begin{itemize}
   \item 
 legal analysis of binding legal sources (GDPR and case-law) and relevant data protection guidelines from the European Data Protection Board and Data Protection Authorities,
document analysis of the IAB Europe TCF, %
    \item 
empirical data gathered on our own website by deploying Quantcast and OneTrust -- the two most popular CMPs in the EU, found respectively on 38.3\% and 16.3\% of the websites with a EU or UK TLD analyzed by Hils et al.~\cite{Hils-etal-20-IMC}.

\end{itemize}
A legal analysis %
is done by a co-author with expertise in Data Protection Law, and a technical analysis by Computer Science co-authors.

In this paper, we make the following \textbf{contributions}: 
\begin{itemize}
    \item we conclude that CMPs process personal data, 
    \item %
we analyse what exact behavior qualifies a CMP as a processor,  
    \item  we identify several scenarios wherein CMPs can qualify as controllers, and
  \item we provide recommendations for policymakers.
\end{itemize}

\section{When are CMPs processing personal data?} 
\label{sec:CMPs}

The \textit{raison d'\^etre} of CMPs is to collect, store, and share a {\em \consentsignal}~\cite{IAB-TCF-v2-TCString, IAB-TCF2020} of a data subject. The \consentsignal\ is a text-based digital representation of the user's consent in a standardised format, stored in the user's browser, and %
provided to third-party vendors by the CMP\cite[paragraph 17, page 9]{IAB-TCF2020}.
Before discussing whether a CMP can be considered a data controller or processor, we first need to establish whether it even falls under the GDPR, which depends on whether it can be considered to process personal data.
To answer this question, we first explain 
the definition of personal data under the GDPR, and then investigate which data CMPs process in practice and whether such data qualifies as personal  data.

\iffalse
According to  the IAB Europe TCF policies~\cite{IAB-TCF2020}, %
CMPs are involved in the processing of (at least) two operations: 
%

i) ‘consent signals’ of a user having consented/rejected, in line with article 5(3) of the ePD and articles 4(11) and 7 of the GDPR. %
\NBtext{IAB Europe says:} \textit{"a CMP must only generate a positive consent Signal on the basis of a clear affirmative action taken by a user"}~\cite[paragraph 3, page 10]{IAB-TCF2020}.  It reflects the options expressed by the user on the consent pop-up which are then sent to the IAB ecosystem;
%
ii) CMPs are also involved in maintaining the records of consent. 
\fi

%
 %
%
%
%
%

%
%
%

%
%
%
%
%
%

%

%

%

\subsection{Legal definitions}
\label{sec:definitions}

\textbf{Personal data} is 
\textit{``any information relating to an identified or identifiable natural person ('data subject'). An identifiable natural person is one who can be identified, directly or indirectly. In particular by reference to an identifier such as a name, an identification number, location data, an online identifier or to one or more factors specific to the physical, physiological, genetic, mental, economic, cultural or social identity of that natural person''} (Article 4(11)~GDPR\cite{GDPR}). 
Recital 30
asserts that online identifiers provided by their devices, such as IP addresses, can be associated to a person, thus making them identifiable. 
 
\noindent
\textbf{Processing} consists of 
\textit{``any operation or set of operations which is performed on personal data or on sets of personal data, whether or not by automated means, such as collection, recording, organisation, structuring, storage, adaptation or alteration, retrieval, consultation, use, disclosure by transmission, dissemination or otherwise making available, alignment or combination, restriction, erasure or destruction''} (Article 4(2) GDPR). In practice, this means that almost any imaginable handling of personal data constitutes processing~\cite{29WP-controller-2010}.
\subsection{Mapping legal definitions into practice }
\label{sec:CMPpersonalData}

{\bf \consentsignal.}
CMPs provide a consent pop-up, encode the user’s choice in a Transparency and Consent (TC) string\footnote{For the sake of uniformity, we call it ``\consentsignal" in the rest of the paper.}, store this value in a user's browser and provide an API for advertisers to access this information. 

IAB Europe TCF specifies that  when \consentsignal\ is "globally-scoped" (shared by CMPs running on different websites), the \consentsignal\ must be stored in a third-party cookie \texttt{euconsent-v2}  set with \verb|.consensu.org| domain. 

CMPs who register at IAB Europe TCF are provided with a subdomain \texttt{<cmp-name>.mgr.consensu.org} that is ``delegated by the Managing Organisation (IAB Europe) to each CMP"~\cite{IAB-TCF-v2-19}.
``Globally-scoped'' \consentsignal\ %
allows all CMPs who manage content on their \texttt{<cmp-name>.mgr.consensu.org} domains to also have access to the \consentsignal\ that is automatically attached to every request sent to any subdomain of \texttt{.consensu.org}. As a result, other consent pop up providers, who are not registered at IAB Europe, are not in a position to receive the \consentsignal\ stored in the user's browser because they have no access to any subdomain of   \texttt{.consensu.org}, owned by IAB Europe. 
For non-global consent, 
a CMP can freely choose which browser storage to use
for \consentsignal~\cite{IAB-TCF-v2-19}. 
%
%
%
%
\iffalse
%

%

\MTcom{The preferences of the user ("consent signal") are encoded in a string of characters.
If the scope of the consent signal is global, it is should be stored as a 3rd party cookie attached to the "consensu.org" domain. Technically, the cookies are stored in the "profile" directory of the user's terminal. More details on cookies: \url{https://developer.mozilla.org/en-US/docs/Web/HTTP/Cookies#define_where_cookies_are_sent}
If it is "service specific", the IAB let the CMP decide: "In version 2 of the TCF Specifications, the storage mechanism used for service-specific TC Strings is up to a CMP, including any non-cookie storage mechanism."
The information on "globally scoped" vs "service specific" consent comes from the TCF specifications on the IAB repo (ref. \cite{IAB-TCF-v2-19} in code -- [34] in pdf).
At all: If you think it is not clear enough, I can rephrase the sentence above.}
\fi 
%
%
%
%
%
%
%
%
%
%
%
%
%
%
The %
\consentsignal\ contains a %
non human-readable encoded version (base64 encoded) of:
\begin{itemize}
    \item
 the list of purposes and features the user consented to;
    \item 
    the list of third-party vendors the user consented for;
    \item 
    the CMP identifier and version, together with %
    other meta-data.
\end{itemize}

%
%
%
%
%
%
%
%
%
%
%
%

\iffalse
\begin{figure}[!tpb]
    \centering
    \includegraphics[width=0.85\textwidth]{fig/ConsentString.png}
    \caption{Storage and distribution of a globally-scoped %
    \consentsignal\ stored in a browser cookie \texttt{euconsent-v2}.}
    \label{fig:ConsentString}
\end{figure}
\fi

%
%
%
%
%

\noindent
{\bf IP address.}
While the %
\consentsignal\ 
does not seem to contain personal data, CMPs additionally have access to the user's IP address. 
In order to include a \consentbanner, publishers are asked to integrate in their website a JavaScript code of a CMP (see step (1) in Figure~\ref{fig:CMPs}). Such code is responsible for the implementation of a \consentbanner\ and in practice is loaded either:
 (1) directly from the server owned by a CMP (OneTrust's banner is loaded from the OneTrust's domain \texttt{https://cmp-cdn.cookielaw.org}), or
 (2) from the server \texttt{<cmp-name>.mgr.consensu.org} 
``delegated by the Managing Organisation (IAB Europe) to each CMP"~\cite{IAB-TCF-v2-19} %
(Quantcast's script %
for \consentbanner\ is loaded from \texttt{https://quantcast.mgr.consensu.org}).
%
 
%

%

%
%

%
\iffalse
Figure~\ref{fig:IP}
provides an overview of two possible ways to load a \consentbanner\ script. 

\begin{figure}[!tpb]
    \centering
    \includegraphics[width=0.85\textwidth]{fig/ConsentString.png}
    \caption{Two ways of loading of the \consentbanner\ script in the user's browser.} 
    \label{fig:IP}
\end{figure}
\fi

%
%
As an inevitable %
consequence of an HTTP(S) request, the server (of a CMP or controlled by a CMP via a DNS delegation by IAB Europe) %
is thus able to access
the IP address of a visitor in this process. 
Additionally, CMP declare in their privacy policies the  collection of IP addresses~\cite{QC-PP,OT-PP}. 
Therefore, from a technical point of view,  a CMP %
is able to record the  IP address of the user's terminal in order to fulfil its service.
Hereby we conclude that CMPs can have access to the user's IP address. 
An IP address can be a cornerstone for data aggregation or identifying individuals.
Empirical studies~\cite{Maier2009,Mish-etal-20-TheWeb} found that a user can, over time, get assigned a set of IP addresses which are unique and stable. Mishra et al.~\cite{Mish-etal-20-TheWeb} found that 87\% of users (out of 2,230 users over a study period of 111 days) retain at least one IP address for more than a month. 2\% of user’s IP addresses did not change for more than 100 days, and 70\% of users %
had at least one IP address constant for more than 2 months. 
These assertions render IP addresses as a relatively reliable and robust way to identify a user.

Even though these results denote IP address  stability (specially static IP addresses), the  data  protection  community  and case law diverge  in  the  understanding of ``dynamic" IP addresses as personal data.
An IP address would be personal data if it relates to an \emph{identified} or \emph{identifiable} person.
It was decided~\cite{Breyer} that a dynamic IP address (temporarily assigned to a device) is not necessarily information related to an \textit{identified} person, due to the fact that ``such an address does not directly reveal the identity of the person who owns the computer from which a website was accessed, or that of another person who might use that computer''.

The question that follows is \textit{whether an IP address relates to an identifiable person for this IP address} to be considered  personal data.
In order to determine whether a person is \textit{identifiable}, account should be taken of \emph{all the means that can reasonably be used} by any entity to identify that person (Recital 26 GDPR). 
This risk-based approach~\cite{Breyer,pallas} means that anyone possessing the means to identify a user, renders such a user identifiable. 
Accordingly, CMPs %
have the means to collect IP addresses (as declared in their privacy policies) and to combine all the information relating to an identifiable person, rendering that combined information (IP address 
and, in some cases, \consentsignal)
personal data. 
Since identifiability of a person depends heavily on context, one should also take into account any other reasonable means CMPs have access to, for example, based on their role and market position in the overall advertising ecosystem~\cite{pallas}. 
One important aspect to consider, then, is the fact that these CMP providers can simultaneously also play a role as an advertising vendor, receiving the Consent Signal provided by their own CMP and (if positive) the personal data of the website visitor. 
Quantcast, for example, appears in the Global Vendor List (GVL)~\cite{IAB-GVL} as registered vendor \#11. In the consent pop-up, their Privacy Policy~\cite{QC-PP}, and their Terms of Service~\cite{QC-Choi-ToS-20,QC-Meas-ToS-20}, Quantcast mentions a large number of purposes for processing personal data, such as ``Create a personalised ads profile'', ``Technically deliver ads or content'', and ``Match and combine offline data sources''. The Evidon Company Directory ~\cite{Evid-QC} labels Quantcast as ``Business Intelligence, Data Aggregator/Supplier, Mobile, Retargeter'', and also mentions a large list of possible personal data collection from them. According to the same source, Quantcast also owns a retargeter called Struq.
In view of this fact, CMPs seem to have reasonable means to combine information relating to an identifiable person, rendering that information personal data.\\

\noindent
{\bf Summary.} Although a \consentsignal\ itself does not seem to contain personal data, 
when the \consentbanner\ script is fetched from a CMP-controlled server,
the CMP also processes the user's IP address, which the GDPR explicitly mentions as personal data. The possibility to combine both types of data renders a user identifiable. 
This possibility becomes particularly pertinent whenever a CMP also plays the role of a data vendor in the advertising ecosystem, which gives them access to more data that could be combined and increase the identifiability of a user.

\section{When are CMPs data processors?}
\label{sec:processors}

\subsection{Legal definitions}

\noindent
A \textbf{processor} is an actor that processes personal data \emph{on behalf} of the controller (Article 4 (8) GDPR). The relevant criteria that define this role are:
 (i) a dependence on the controller's instructions regarding processing activities%
 ~\cite{29WP-controller-2010}, (Art. 28(3)(a)), Recital 81), and;
 (ii) a compliance with those instructions  %
 ~\cite{EDPB-controller-2020}, which means they are not allowed to go beyond what they are asked to do by the controller\cite{EDPB-controller-2020}.

\subsection{Mapping legal definitions into practice}
The main objectives of CMPs clearly correspond to the definition of 
 data processors, because they act according to the instructions given by the website publisher with regards to the legal bases, purposes, special features, and/or vendors to show to the user in the consent pop-up. IAB Europe TCF also explicitly defines CMPs as data processors in the TCF documentation~\cite[page 10 (paragraph 8), page 11 (paragraph 11)]{IAB-TCF2020}. 
 The classification of the CMP as data processors is currently the widely shared consensus about their role. \\
%
%
%
%

%
%
%

\iffalse
\begin{itemize}
    \item CMPs \emph{``may be instructed by its Publisher which Purposes, Special Features, and/or Vendors to disclose''}~\cite[paragraph 8, page 10]{IAB-TCF2020}.
    
     \item CMPs \emph{``may be instructed by its Publisher to establish, record and transmit information about its own Legal Bases''}~\cite[page 11, paragraph 11]{IAB-TCF2020}.
     
%
 
     %
 %
\end{itemize}
\fi

%
%
%
%

%

%

%

%

%

%

\iffalse
While fulfilling its functions, the access to personal data is generally not the main object or target of a CMP.
However, due to the architecture imposed by IAB Europe to maintain records of globally-scoped consent signals on a server managed by IAB Europe and delegated to CMPs, \verb|*.mgr.consensu.org|~\cite{IAB-TCF-v2-19}, it is inevitable that the CMP might entail access to personal data when performing the service.\\ 
\fi 
%
%
%
%

\noindent
\textbf{Responsibility of CMPs  as processors.}
If a CMP is established as a data processor, it can be held liable and fined if it fails to comply with its obligations under the GDPR (Articles 28(3)(f) and 32-36 GDPR).
Moreover, if a false %
\consentsignal\ 
is stored and transmitted, it may well be %
considered an
``unauthorised disclosure of, or access to personal data transmitted, stored or otherwise processed''
~\cite[Art. 32(2)]{GDPR}. %

Recent works reported numerous CMPs violating the legal requirements for a valid positive consent signal under the GDPR. For example, researchers detected pre-ticked boxes~\cite{Nouwens-etal-20-CHI,Matt-etal-20-IEEESP}, refusal being harder than acceptance~\cite{Nouwens-etal-20-CHI} or not possible at all~\cite{Matt-etal-20-IEEESP}, choices of users not being respected~\cite{Matt-etal-20-IEEESP}, as well as more fine-grained configuration barriers such as aesthetic manipulation~\cite[Fig. 11]{Gray-etal-21-CHI-2}, framing and false hierarchy~\cite[Fig. 12]{Gray-etal-21-CHI-2}.

\section{When are CMPs data controllers?}  
\label{sec:controllers}

In this section we analyse when CMPs are data controllers. 
Firstly, in section \ref{sec:controllers_definitions} we provide the legal definitions necessary to qualify CMPs as data controllers.

In the following sections (\ref{sec:tracking} -- \ref{sec:dark-patterns})
we will map these legal definitions into practice. 
Although CMPs are explicitly designated as processors by the IAB Europe TCF specifications~\cite{IAB-TCF2020}, we analyse four functional activities of CMPs that enables their qualification as data controllers. We include a technical description of such activities followed by a legal analysis. These activities refer to:

\begin{quote}

\begin{itemize}
    \item [\S \ref{sec:tracking}] Including additional processing activities in their tools beyond those specified by the IAB Europe;
    \item [\S \ref{sec:scanning}] Scanning publisher websites for tracking technologies and sorting them into 
    purpose categories;
    \iffull
    \item [\S \ref{sec:vendors}] Controlling third-party vendors included by CMPs;
    \fi
    \item [\S \ref{sec:dark-patterns}] Deploying manipulative design strategies in %
    the UI of consent pop-ups.
\end{itemize}

\end{quote}

Finally, in section \ref{sec:liability}  we determine the responsibility of a CMPs as data controllers.

\subsection{Legal definitions}\label{sec:controllers_definitions}

\noindent
The primary factor defining a \textbf{controller} is that it 
``determines the purposes and means of the processing of personal data'' (Article 4(7) GDPR).
We refer to the European Data Protection Board (EDPB) opinion~\cite{29WP-controller-2010} to unpack what is meant by 1) ``determines'', and 2) ``purposes and means of the processing of personal data''.

\noindent
\textbf{``Determines'} refers to having the ``determinative influence'', ``decision-making power''~\cite{EDPB-controller-2020,29WP-controller-2010, CaseC-25/17} or ``independent control''~\cite{ico-guidance-controllers} over the purposes and means of the processing.  %
This concept of ``determination'' provides some degree of flexibility (to be adapted to complex environments) and the Court of Justice of the EU (CJEU), Data Protection Authorities (DPAs) and the EDPB describe that such control can be derived from: 
\begin{itemize}
    \item professional competence (legal or implicit)~\cite{29WP-controller-2010};
    \item factual influence %
based on %
factual circumstances surrounding the processing.
(e.g. to contracts, and real interactions)~\cite{29WP-controller-2010};
\item image given to
data subjects and their reasonable expectations on the basis of this visibility~\cite{29WP-controller-2010}; 

\item which actor “\emph{organizes, coordinates and encourages}” data processing~\cite{CaseC-25/17} (paragraphs 70, 71);

\item  interpretation or independent
judgement exercised to perform a professional service~\cite{ico-guidance-controllers}.

\end{itemize}

\noindent
\textbf{ ``Purposes'' and ``means''} refer to ``why'' data is processed (purposes) and ``how'' the objectives of processing are achieved (means).
Regarding the determination of ``purposes", the GDPR merely refers that purposes need to be explicit, specified and legitimate (Article 5(1)(b)~\cite{Foua-etal-20-IWPE}. 
In relation to the determination of "means", the EDPB distinguishes between ``essential'' and ``non-essential
means'' and provides examples thereof~\cite{EDPB-controller-2020, 29WP-controller-2010}:
\begin{itemize}
    \item ``Essential means'' are inherently reserved to the controller; %
    examples are: determining the i) type of personal data processed, ii) duration of processing, iii) recipients, and iv) categories of data subjects;
    
    \item ``Non-essential means'' may be delegated to the processor to decide upon, and concern the practical aspects of implementation, such as: i) choice for a particular type of hardware or software, ii) security measures, iii) methods to store or retrieve data.
\end{itemize}
\iffull
Important notes on the assessment of controllers are referred herewith. The role of controller and processor are \emph{functional} concepts~\cite{EDPB-controller-2020}: the designation of an actor as one or the other is derived from their \emph{factual roles and activities} in a specific situation~\cite{29WP-controller-2010}, rather than from their formal designation~\cite{Opinion-Adv-Mengozzi}. 
Notably, access to personal data is not a necessary condition to be a controller~\cite{CaseC210/16,CaseC40/17}.
Moreover, the control exercised by a data controller may extend to the entirety of processing at issue, and also be limited to {\em a particular stage in the processing}~\cite{CaseC40/17}.
\fi

\subsection{Inclusion of additional  processing  activities}
\label{sec:tracking}

\noindent
{\bf Technical description.}
When publishers employ the services of a CMP to manage consent on their website, the CMP provides the publisher with the necessary code to add their consent solution to the website.
Although this code is ostensibly only for managing consent, it is possible for the CMP to also include other functionality. 

As part of our empirical data gathering, we assumed the role of website owner (i.e., publisher) and installed a QuantCast CMP~\cite{QC-Choice} on an empty website.
Website owners are instructed by the CMP to ``copy and paste the full tag'' into their website header and ``avoid modifying the tag as changes may prevent the CMP from working properly.''~\cite{QC-Tag}: the tag is the minimal amount of code necessary to load the rest of the consent management platform from an external source.

\iffull
\begin{figure}[!tpb]
    \centering
    \includegraphics[width=0.85\textwidth]{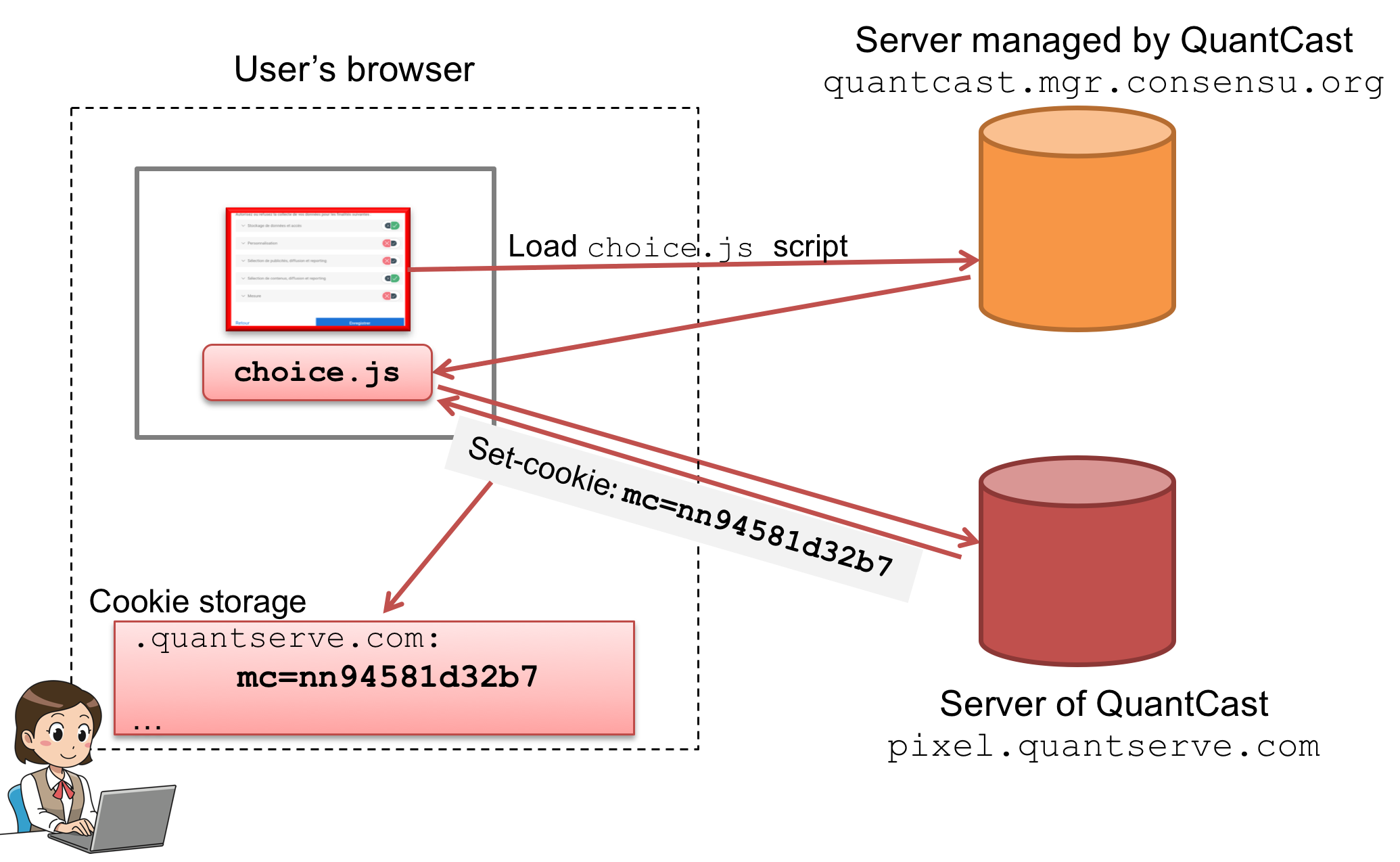}
    \caption{Loading of invisible pixel by a QuantCast consent pop-up: the pixel sets a third-party  cookie \texttt{mc} with a user-specific identifier that is further accessible to all subdomains of \texttt{quantserve.com}.}
    \label{fig:Pixel}
\end{figure}
\fi

When installing the Quantcast CMP, we discovered that the ``Quantcast Tag'' script that deploys a consent pop-up on the website also loads a further script \texttt{choice.js} that integrates a 1x1 invisible image loaded from the domain  \texttt{pixel.quantserve.com}~\iffull (see Figure~\ref{fig:Pixel})\fi.
When this image is loaded, it also sets a third-party cookie \texttt{mc} in the user's browser.
By replicating the methodology to detect trackers~\cite{Foua-etal-20-PETs}, we 
analysed %
the \texttt{mc} cookie from \texttt{pixel.quantserve.com};
this cookie is {\em ``user-specific''} -- that is, its value is different for different website visitors -- and comes from a third-party, allowing tracking across all sites where some content from \texttt{quantserve.com} or its subdomains is present. 
Such tracking by \texttt{quantserve.com} is prevalent in practice: recent research shows that third-party trackers from  {QuantCast} are in top-10 tracking domains included by other trackers on 9K most popular websites~\cite[Fig. 6]{Foua-etal-20-PETs}.

In the documentation that describes the QuantCast CMP, they mention that their CMP also contains a ``QuantCast Measure'' product~\cite{QC-Tag} that is labeled as {\em ``audience, insight and analytics tool''} for {\em ``better understanding of audience''}~\cite{QC-Measure}.
The \verb|mc| cookie we detected is the only cookie present on our empty website {\em before interacting with the QuantCast pop-up}, and thus we conclude that this cookie is likely responsible for the audience measurement purpose of QuantCast. \\

\noindent
{\bf Legal analysis.} 
The QuantCast script installs {\em both a consent pop-up and a tracking cookie}, and its technical implementation makes it impossible for website owners to split these two functionalities. %
Such joint functionality triggers  consequences on its legal status. 
The tracking cookie enables the QuantCast CMP to process data for its own tracking and measurement purposes, regardless of any instructions from the publisher, nor from the specifications of the IAB Europe TCF.
Hence, the  independent and determinative influence of a CMP is based on factual  circumstances  surrounding   the   processing, which qualifies a CMP {in this scenario} as a data controller. 

\subsection{Scanning and pre-sorting of tracking technologies} 
\label{sec:scanning}

\noindent
{\bf Technical description.} 
One of the services CMPs often provide to publishers is a \emph{scanning technology} which identifies the tracking technologies currently installed and active on the publisher's website (e.g., ``first- and third-party cookies, tags, trackers, pixels, beacons and more''~ \cite{cookieproscan}).
This scan is generally the first step when installing a consent pop-up on the website, and can be configured to automatically repeat on a regular basis.

In addition to providing descriptive statistics on the trackers currently active (e.g., what type of tracking), the scan results also include a \emph{pre-sorting} of each of these technologies {\em into a particular data processing category}  which are then displayed in the banner.
In the case of OneTrust's CookiePro scanner, which is integrated into the banner configuration procedure when it is performed with an account, trackers are \emph{``assigned a Category based on information in the Cookiepedia database''}~\cite{cookieprolesson3,cookiepedia} (a service operated by OneTrust itself).
The scanning includes identifying trackers (and matching them with vendors using %
Cookiepedia) and categorising these trackers/vendors in specific purposes.
The four common purposes of trackers of Cookiepedia are i) strictly necessary (which includes authentication and user-security); ii) performance (also known as analytics, statistics or measurement); iii) functionality (includes customization, multimedia content, and social media plugin); and iv) targeting (known as advertising).
Any trackers which cannot be found in the database are categorised as ``Unknown'' and require manual sorting (see Figure~\ref{fig:cookiepro_scanresults}%
). 
From the setup guides, there seems to be no explicit or granular confirmation required by the publisher itself (although they can edit after the fact): once the scan is complete, the categorisation of trackers is performed automatically and the consent pop-up is updated.
In other words, the CookiePro's \consentbanner\ interface is in part automatically configured by the scanning tool.

This kind of scanning and categorising feature based on a CMPs own database is also offered by several other CMPs such as Cookiebot~\cite{cookiebotscanner}, Crownpeak~\cite{crownpeakcategories}, TrustArc~\cite{trustarcscanner} and Signatu~\cite{signatuscanner}.\\

\begin{figure}[!tpb]
    \centering
    \includegraphics[width=0.95\textwidth]{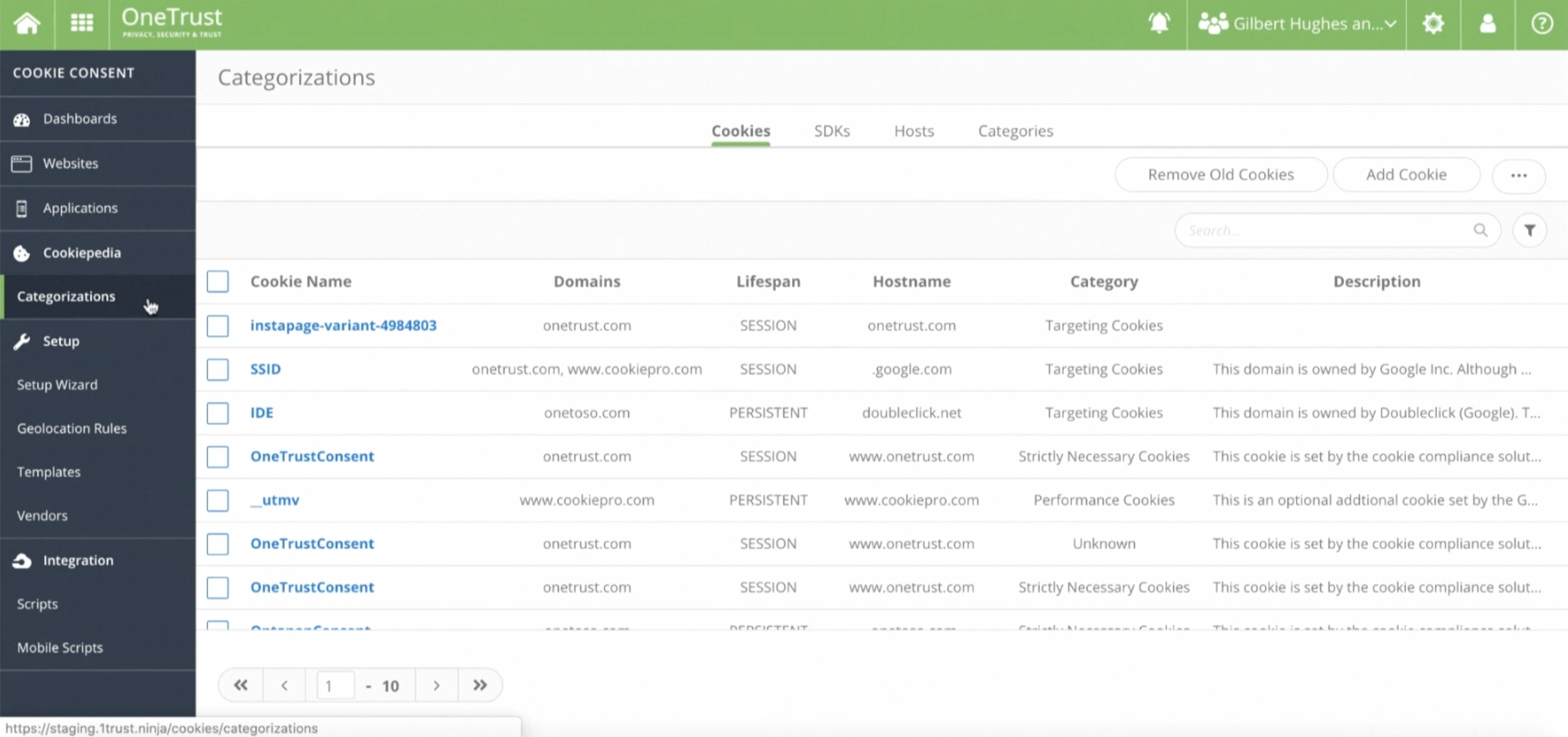}
    \caption{CookiePro's configuration back-end designed for the publisher, when logged. After completing a scan for trackers on the publisher's website, this screen shows the trackers that were found
    together with a category they are assigned with.}
    \label{fig:cookiepro_scanresults}
\end{figure}

\noindent
{\bf Legal analysis.}
In this concrete scenario, through providing the additional services and tooling (besides consent management) of scanning and consequently presorting tracking technologies into pre-defined purposes of data processing, CMPs contribute to the definition of purposes and to the overall compliance of the publisher wherein the CMP is integrated. 
This level of control of a CMP in determining the purposes for processing personal data and means is a decisive factor to their legal status as data controllers. 

Moreover, CMPs that offer this additional service can be potentially be qualified as a \emph{joint controller} (Article 26 GDPR) together with the publisher, as both actors jointly determine the purposes and means of processing. 
In line with the criteria provided by the EDPB~\cite{EDPB-controller-2020}, these additional processing operations convey the factual indication of a pluralistic control on the determination of purposes from this concrete CMP and respective publisher embedding these services by default. 
The acceptance of scanning and categorization of purposes entails  i) a \emph{common and complementing} decision taken by both entities, wherein the categorization of purposes ii) is \emph{necessary} for the processing to take
place in such manner that it has a \emph{tangible impact} on the determination of the purposes and
means of the processing and on the overall and forthcoming data processing.

The provision of both %
\consentbanner\ and scanning tool services 
by a CMP to a publisher creates a situation of \emph{mutual
benefit}~\cite{CaseC40/17, CaseC210/16}: CMPs provide a service that creates a competitive advantage compared to other CMP providers, and publishers are relieved of having to manually match trackers with vendors, purposes, and legal bases.

As joint controllers, both entities  would then need to make a transparent agreement %
to determine and agree on their respective responsibilities for compliance with the obligations and principles under the GDPR, considering also the exercise of data subjects’ rights and the duties to provide information  as required by Articles 13 and 14 of the GDPR. The essence of such arrangement must be made available to the data subject~\cite{EDPB-controller-2020}.

Such joint responsibility does not necessarily imply equal responsibility of both operators~\cite{CaseC210/16}, nor does it need to cover all processing, in other words, it may be limited to this particular stage in the processing of scanning and presorting of trackers~\cite{CaseC40/17}.

\iffull
\subsection{Controlling third-party vendors included by CMPs} 
\label{sec:vendors}

\noindent
{\bf Technical description.}
Upon installation of a CMP, the website publisher generally has the possibility to decide which vendors (third-party advertisers) to include in the \consentbanner. 
From more than 600 vendors currently registered at IAB Europe TCF~\cite{IAB-GVL}, only the selected vendors will be then stored in the %
\consentsignal\ 
when the user interacts with the \consentbanner. 
In practice, the way the publisher effectively exercises this choice of vendors depends on the options available in the configuration tool provided by the CMP.

The IAB policies explicitly state that a CMP cannot have preferential treatment for one vendor or another~\cite[paragraph 6(3)]{IAB-TCF2020}. 
\iffalse
\begin{quote}
  \emph{``In any interaction with the Framework, a CMP may not exclude, discriminate against, or give preferential treatment to a Vendor except pursuant to explicit instructions from the Publisher involved in that interaction and in accordance with the Specifications and the Policies. For the avoidance of doubt, nothing in this paragraph prevents a private CMP from fully implementing instructions from its Publisher owner.''}
\end{quote}
\fi
Hence, CMPs cannot pre-select or treat vendors differently, unless {\em a publisher explicitly asks} a CMP to include/delete some vendors from the list of all vendors.

Herewith we analyse two case studies of QuantCast and OneTrust.
Figure~\ref{fig:vendors} shows an installation process of 
QuantCast CMP, which gives some power to  publishers.
It includes by default  {\em around 671 vendors registered in the IAB Europe TCF}, but allows a publisher to remove some of the vendors from this list. 
This power given to publishers is, however, limited: publishers must either manually search and {\em select one-by-one the vendors they want to exclude}. 
%
\iffalse
\MTcom{Technically, they can also block the whole list of vendors, but for this they need to tick the checkbox on the top of the list (like in a mailbox) and then click on "block selected vendors". Then, they will be able to re-enable vendors one-by-one. I think we should precise this point. What do you think?}
\NBcom{it's very interesting and if you can write it in 2 sentences, then I'd say: do it ;-)}
\MTcom{I tried to rephrase the sentences above to add this information.}
\NBcom{Sincerely, it's hard to understand. Better to remove these details IMO. I commented out your new text above.}
\fi

%
%
%
\begin{figure}[htb]
  \centering
    \includegraphics[width=0.95\textwidth]{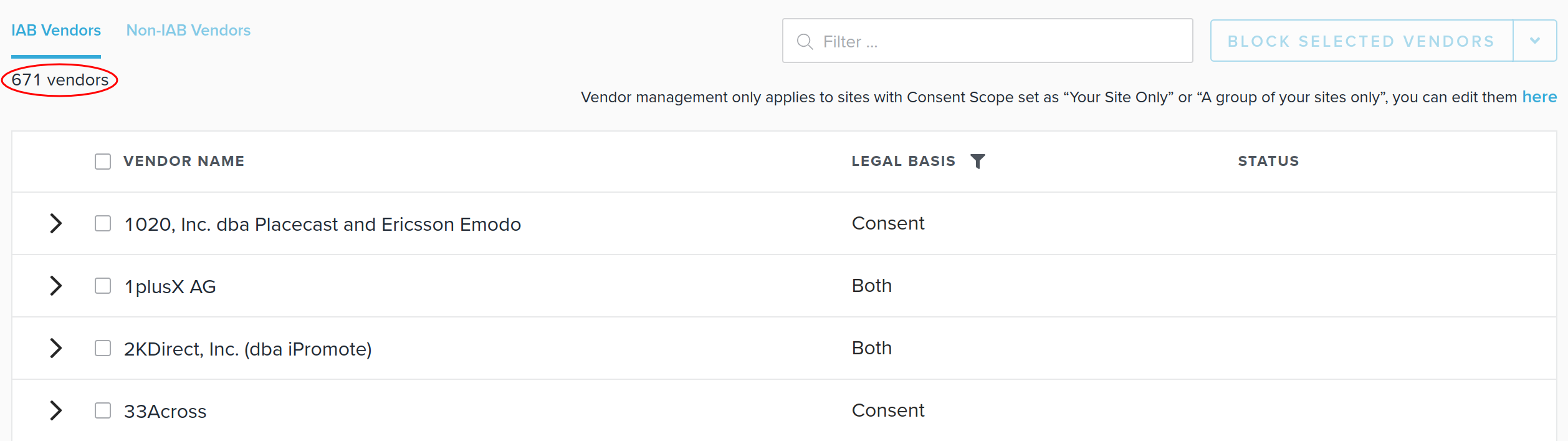}
    \caption{Installation process of QuantCast CMP [Captured on 5 Feb. 2021].  A publisher has to manually search and { exclude one-by-one the vendors from the list of 671 vendors registered in  IAB Europe TCF.}}
  \label{fig:vendors}
\end{figure}
Regarding OneTrust's free, open access service called \textit{CookiePro Free IAB TCF 2.0 CMP Builder}~\cite{cookiepro-tcf-2-builder}, it gives no control to the publisher over the list of vendors to include.
As a result, when the user clicks ``Accept'' in a CookiePro banner we installed on our empty website, the \consentsignal\  contains 2 special features optin, 10 purposes under the legal basis of consent, 9 purposes under legitimate interest, \emph{631 vendors for consent}, and 261 vendors under %
legitimate interest.

Relying on a publisher to manually remove the vendors with whom it does not have a partnership presupposes that publishers are willing to actively check and configure the list of vendors, which can require an active action from a publisher on a separate screen during the configuration process.
Such assumption contends with relevant findings from behavioral studies regarding \emph{default effect bias}, referring to the tendency to stick to default options~\cite{John-etal-02-ML,John-Gold-03-Science,Spoo-UserSettings-11,NCC-Deceived-18}.
Thaler and Sunstein concluded that \textit{``many people will take whatever option requires the least effort, or the path of least resistance''}~\cite{Thal-Sust-2019}.
It seems reasonable to argue that publishers will generally leave the list as is. 
\\

\noindent
{\bf Legal analysis.}
 CMPs are in a position to decide what decision-making power to award to website publishers regarding the selection of specific vendors. 
 By restricting the ability of the publisher to (de)select vendors, the CMP obliges the publisher to present to the user the full list of IAB Europe-registered vendors. 
We recall that  when registering to the IAB Europe, each vendor declares a number of purposes upon which it wishes to operate, and hence it can be concluded that the CMP automatically increases the number of purposes displayed to -- and possibly accepted by --  the end-user. 
As a result, a CMP requires the publisher to present more processing purposes than necessary, which has direct consequences on the interface the end-user will interact with. 

With such factual decision-making power over the display of purposes rendered to users, it can be observed that CMPs exert influence over the determination of purposes of data processing, turning it to a data controller. 
Relatedly, deciding on the third-parties that process personal data consists on the determination of \emph{"essential means"} -- a competency allocated only to controllers, which again consolidates our conclusion that CMPs are data controllers in the above mentioned scenario.

This practice of including by default hundreds of third-party vendors implies that CMPs seem to breach several data protection principles:

\noindent
{\em Transparency and fairness principle} (Article 5(1)(a) GDPR) which mandates  controllers to handle data in a way that would be reasonably expected by the data subjects. When users signify their preferences in the consent pop-up, they are not aware nor expect their data to be potentially shared with around 600 third-parties. 
Moreover, the inclusiveness by default of this amount of partners seems to trigger severe risks to the rights of users and thus this consent sharing needs to be limited (Recital 75 GDPR). 

\noindent
{\em Minimization principle} (Article 5(1)(c) GDPR) provides that data shall be "adequate, relevant and limited to what is necessary in relation  to the purposes for which they are processes". This principle is generally interpreted as referring to the need to minimise the quantity of data that is processed. One may, however, also wonder whether the principle extends to other characteristics such as the number of recipients to which data is shared with. 
Moreover, according to the theory of choice proliferation, a large number of purposes can lead to the user experiencing negative effects. However, in the case of \consentbanner s, the critical threshold of presented purposes beyond which these effects occur is not yet known~\cite{Mach-Bohm-20-PoPETS}.

\fi

\subsection{Deployment of manipulative design strategies} 
\label{sec:dark-patterns}

\noindent
{\bf Legal compliance vs. consent rates.}
When designing their consent pop-ups, CMPs have considerable freedom:
The only constraint placed on them by the IAB's TCF is that they need to include the purposes and features exactly as defined by the IAB Europe~\cite{IAB-TCF2020}. 
From a UI perspective, CMPs thus enjoy a design space and can choose {\em how exactly these choices are presented to the end user}. 

The primary service offered by CMPs is to ensure legal compliance, which largely determines how they exercise their design freedom. 
However, the advertising industry is also incentivised to strive for \textit{maximum consent rates}.
This is apparent when looking at how CMPs market themselves. For example, Quantcast describes their tool as able to ``\emph{Protect and maximize ad revenue while supporting compliance with data protection laws}''~\cite{QC-Choice} and provides ``Choice Reports'' that detail ``[h]ow many times Choice was shown, Consent rate and Bounce Rate and a detailed breakout if the full, partial or no consent given''~\cite{QC-user-guide}.
OneTrust advertises that its CMP can ``\emph{optimize consent rates while ensuring compliance}'', and ``\emph{leverage A/B testing to maximize engagement, opt-ins and ad revenue}''~\cite{OneTrustCMP}.
In other words, although the official and primary service provided by CMPs is legal compliance, in practice, their service consists in {\em finding the balance between strict legal compliance and maximum consent rates} (considered to be negatively correlated), and this balancing ability becomes a point of competition between them. 
\noindent
{\bf Manipulative design strategies in \consentbanner s.}
Recent works denote that
many popular CMPs deploy manipulative design strategies in \consentbanner s~\cite{Nouwens-etal-20-CHI,Matt-etal-20-IEEESP,Gray-etal-21-CHI-2} and that such strategies influence the users' consent decisions~\cite{Nouwens-etal-20-CHI, Utz-etal-19-CCS}.
In concrete, recent findings concernedly report the majority of users think that a website cannot be used without giving consent  (declining trackers would prevent access to the website) and also click the "accept" button of the banner out of habit~\cite{Utz-etal-19-CCS}.

\noindent
{\bf Technical analysis of default \consentbanner s. }
We portray an illustrative example of the use of manipulative design strategies in a consent pop-up. We installed a free version of OneTrust \consentbanner, the \textit{CookiePro Free IAB TCF 2.0 CMP Builder}, on our empty website. During the installation, we chose a default version of the banner without any customization. 
Figure~\ref{fig:cookiepro-banner}
depicts the 2\textsuperscript{nd} layer of the CookiePro's default banner: the option to ``Accept All'' is presented on top of the banner, (hence making acceptance to all purposes prioritized), 
while ``Reject All'' and ``Confirm My Choices'' are located at the very bottom of the banner, only made available after scrolling down. This banner includes the dark patterns of ``obstruction'', ``false hierarchy'' and ``sneaking''~\cite{Gray-etal-18-CHI}. 

\begin{figure}[tb!]
  \centering
    \includegraphics[width=0.95\textwidth]{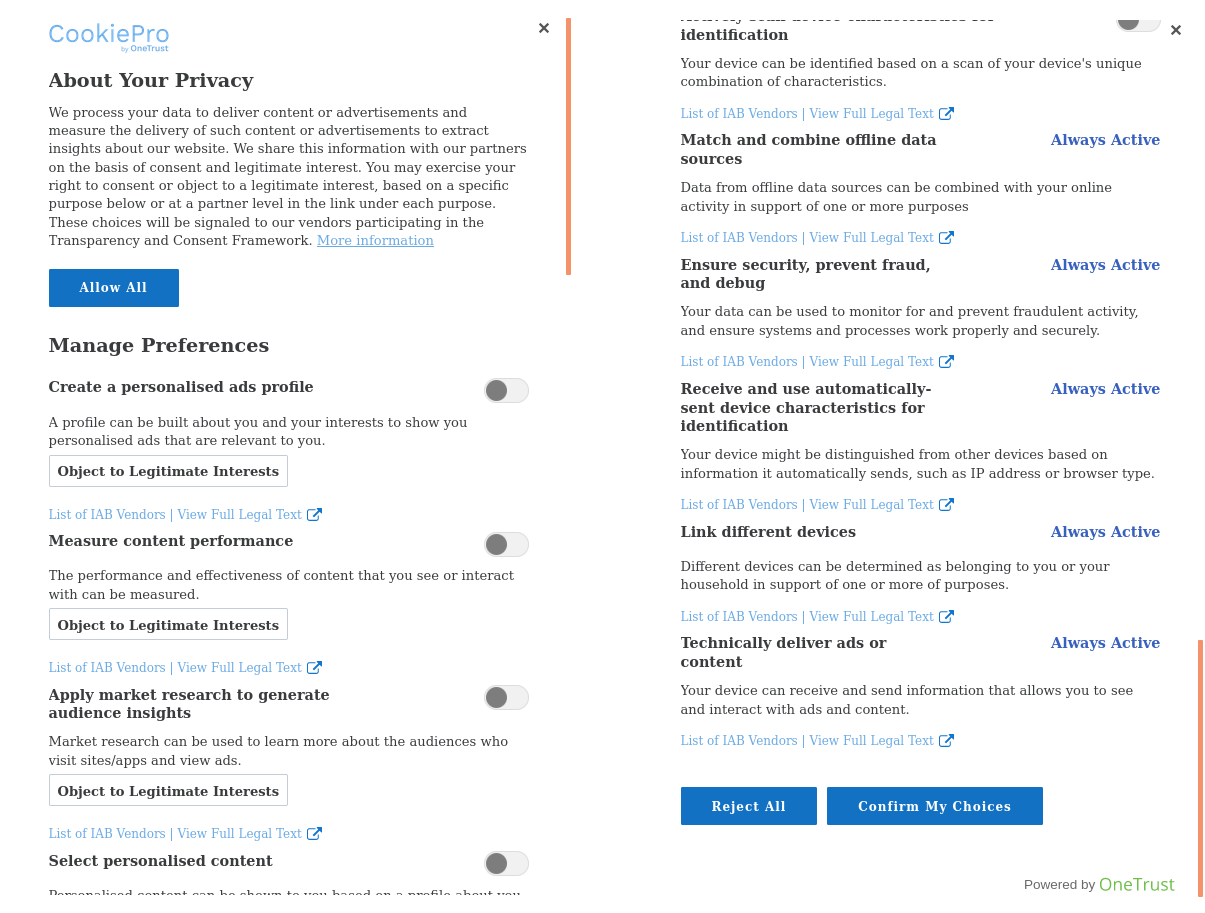}
    \caption{2\textsuperscript{nd} layer of the default consent pop-up provided by CookiePro Free IAB TCF 2.0 CMP Builder (owned by OneTrust). [Captured on 13 Jan. 2021]. On the left, the top level of the page, displaying the ``Accept All'' button. On the right, the bottom of the same screen, displaying the ``Reject All'' and ``Confirm My Choices'' buttons, so the user needs to scroll down in order to see them.}
  \label{fig:cookiepro-banner}
\end{figure}
%

%
%

%
%
%
%
\iffalse
This holds particularly relevant where the most important options offered (purposes, accept/reject options) should be given in parallel, in equal footing (in terms of format, position, color, tone, design, size) rather than hierarchical~\cite{Sant-etal-20-TechReg}. 
%
Consequently, this %
design convinces the user to select what they feel is either the only option (presented on top), or the best option (proposed in a better position), while other options (to refuse) are cumbersome and hidden.
\fi 
%

\noindent
{\bf Legal analysis.}
From a regulatory perspective, several guidelines  have  been  issued  by the EU Data  Protection Authorities  on  \consentbanner s, suggesting UI should be designed to ensure that {\em user's choices are not affected by interface designs}, proposing a privacy by design and by default approach (Article 25 GDPR), wherein default setting must be designed with data protection in mind. Proposals of such design refer that options of the same size, tone, position and color ought to be used, so as to provide the same level of reception to the attention of the user)~\cite{CNIL-Shap-19,CNIL-Guid-20,Greek-Guid-20,ICO-Guid-19,Irish-Guid-20,AEPD-Guid-21}.
Although these guidelines are welcomed, they do not have enough legal power to be enforceable in court, and it is unclear whether they impact compliance rates. 
However, in practice a CookiePro default design 
convinces the user to select what they feel is either the only option (presented on top), or the best option (proposed in a better position), while other options (to refuse) are cumbersome and hidden.

\noindent
{\bf Determination of means.}
The primary service of CMPs is to provide consent management solutions to publishers through consent pop-ups, and thus anything related to this service can be considered as part of the ``non-essential means'' that can be delegated to a processor (see Section~\ref{sec:controllers_definitions}). However, when CMPs decide to include manipulative design strategies -- known as {\em dark patterns} -- to increase consent optimization rate, these can be considered to go beyond their primary goal.
Manipulating users decision-making to increase the probability of prompt agreement to consent for tracking is not strictly necessary to provide its consent management service.
In particular, resorting to such interface design strategies does not seem to consist of "basic features" or "service improvement" that could be considered as normally expected or compatible within the range of a processor's services~\cite{Hintze2018DataCD}. 
In fact, there are no technical reasons that could substantiate the recourse to these dark patterns. A CMP could devise design banners in a fair and transparent way and which complies with the GDPR. 
The EDPB~\cite{EDPB-consent-2020} %
refers that \textit{"compulsion to agree with the use of personal data additional to what is strictly necessary, limits data
subject’s choices and stands in the way of free consent."}
We conclude that the use of manipulative strategies does not qualify as a mere technical implementation or operational use to obtain lawful consent, and instead falls inside the ``\emph{essential means}'' category, making them a data controller. 

\noindent
{\bf Determination of purposes.}
Following the cognition held by the CJEU on the Jehowa’s Witnesses case~\cite{CaseC-25/17},
one decisive factor of the role of a controller consists in the determination of “\emph{who organized, coordinated and encouraged}” the data processing (paragraphs 70, 71).
CMPs have exclusive \emph{judgement and control} to
 adopt manipulative design strategies. Such strategies have
 a real impact on users' consent decisions and ultimately impact the processing of their data.
By deploying such strategies, CMPs do not act on behalf of any other actor (which would lead to them being recognized as ``processors''), but instead have control over which purposes will be more likely to be accepted or rejected by users. 
In practice, CMPs' deployment of {\em dark patterns} that manipulate the user's final choice
evidences a degree of \emph{factual influence or decision-making power} 
over the processing activities that will follow. %

\noindent
{\bf Summary.} 
CMPs exercise a dominant role in the decision-making power on eventual processing activities within the IAB Europe TCF ecosystem.
We %
argue that whenever CMPs impose dark patterns to a publisher and similarly whenever CMPs propose a default banner that features dark patterns to a publisher,
these facts strongly indicate a controllership status in its own right due to CMPs' influence on the determination of means and purposes of processing, even if only to a limited extent. 
However, the afforded discretion availed to CMPs requires a case by case analysis and is more likely to lead to divergent interpretations.

\iffull
\subsection{What is the responsibility of a CMP as controller?}
\label{sec:liability}

A CMP as a data processor that goes beyond the mandate given by the controller and acquires a relevant role in determining its own purposes, as shown in the %
scenarios in Section~\ref{sec:controllers},
becomes a controller with regard to those specific processing operations~\cite{29WP-controller-2010} and will be in breach of its obligations, hence subject to sanctions (Article 28(10)). 
The breadth of the parties responsibility, including the extent to which they become data controllers, should be analysed on a case by case basis~\cite{WP29-advertising} depending on the particular conditions of collaboration between publishers and CMPs, and then should be reflected in the service agreements. 

One of their responsibilities as controllers include the obligation to comply with the principles of data protection, thereby they are required to obtain personal data fairly, lawfully and to  comply with any transparency requirements with respect to users and obtain a valid consent.

Additionally, CMPs should offer design choices that are the most privacy-friendly, in a clear manner and as a default choice, in line with the principle of data protection by design and data protection by default  (Article 25 of the GDPR).
Finally, CMPs should respect the minimization principle -- the use of compulsion  methods (either in the manipulation of purposes, either pre-registering around 600 vendors) \textit{to agree with the use of personal data additional to what is strictly necessary limits data subject’s choices and stands in the way of free consent}~\cite[paragraph 27]{EDPB-consent-2020}.

\section{Recommendations}

In this section, based on our legal and empirical analysis, we propose a number of recommendations for policy makers that could address the current ambiguity revolving the role of CMPs.\\

\noindent
{\bf Concepts of controller and processor in the GDPR  need to be clarified.}
We hope to provide influential stakeholders, such as the EDPB, with operational information that can inform its next guidelines on the concepts of controller and processor in the GDPR~\cite{EDPB-controller-2020}. In particular, and in the context of the current paper, we would recommend to %
clarify the following aspects:
\begin{enumerate}
    \item on defining purposes in practice: our work shows that a CMP influencing users decision-making with respect to accepting or rejecting pre-defined purposes actually renders such entity co-responsible for determining purposes;
  \item on the role of deploying manipulative design in CMPs and whether this constitutes “essential means” of processing; 

    \item on the contractual agreement between publishers and CMPs: such agreement  should mirror as much as possible the factual roles and activities they are involved in, pursuant to legal certainty and transparency;

\end{enumerate} 

\noindent
{\bf Guidelines needed on ``provision of services" for data processors.}
 Data processors must limit its operations to carrying out the services for which
the controller stipulated in the processing agreement. However, this design space is left to ambiguity and leeway in terms of what “providing the service” entails. Guidance is needed on what is considered to be \emph{compatible and expected purposes} for the provision of their services/operations. For example, while security operations are surely expected, doubts remain regarding the provision of services which include other purposes that go beyond legal provisions and principles such as the compatibility between optimization of consent rate and legal compliance (as mentioned in Section~\ref{sec:dark-patterns}); the EDPB~\cite[paragraph 27]{EDPB-consent-2020} mentions that such goal cannot be prioritized over the control of an individual's personal data:
\textit{ %
an individual’s control over their personal data is essential and there is a strong presumption that consent to the processing of personal data that is unnecessary, cannot be seen as a mandatory consideration in exchange for the performance of a contract or the provision of a service}. \\

\noindent
{\bf DPAs should scale up %
auditing of CMPs. }
Currently, DPAs primarily use labour-intensive, small-sample, qualitative methods to evaluate the legal compliance of CMPs (e.g., 
the Irish DPA analysed \consentbanner s of 38 websites via 
a ``desktop examination'' 
\cite{DPC-report}). 
Although our normative stance is that compliance evaluations should not be outsourced to algorithms and always involve human oversight, data-driven and automated tools could help DPAs gain a broader understanding of CMP design and compliance trends within their jurisdiction.
Auditing can be automated (for example, with scraping technologies)
to analyse the presence or absence of certain consent options (e.g., a reject button), interaction flows (e.g., number of clicks to access an option), or default settings (e.g., checked or unchecked choices). 
Not all requirements for consent are as binary and can be measured in this way (such as the quality of purpose descriptions), but gathering and continuously monitoring those aspects %
can provide DPAs with initial indications.
These insights can be used to decide which follow-up investigations are necessary, and also which aspects might provide the biggest impact if addressed.\\

\noindent
{\bf Automated auditing of CMPs requires extension of consent signal.}
The IAB Europe has created a standardised format for consent signals and successfully implemented APIs that allow 
various entities to interoperate with each other.
Such consent was created to simplify the exchange of the digital version of consent between CMPs and advertisers. They do not, however, contain elements that could help DPAs and users to evaluate the \emph{validity of collected consent} through automated means.
We strongly suggest these standards and APIs should be expanded (or new ones developed by neutral parties) 
to include information about the interface design of a \consentbanner.
Such extended  digital format of consent will make consent services
computationally legible by more actors, such as regulators and researchers.

Additionally, in the current IAB Europe TCF system, third-party advertisers (vendors) just receive a \consentsignal\ as a part of HTTP(S) request or via browser APIs, but there is no proof whether such \consentsignal\ is valid and whether a vendor actually received it (or, for example, did not generate it by itself instead). We recommend IAB Europe TCF to change this practice and to propose solutions that demonstrate evidence of consent collection and its integrity.\\

\noindent
{\bf Guidance needed on validity of pre-registration of vendors.} 
Through our analysis, we identified that CMPs have the capability to ``pre-register'' about 600 vendors during the installation process on a website. This pre-registration of vendors means that if the user accepts some of the purposes presented in the consent pop up, then all the vendors will be automatically added to a %
\consentsignal\ 
(see an example of OneTrust in section~\ref{sec:vendors}, where 632 vendors are allowed when the user clicks ``Accept''). 
Consent stored by CMP in this case {\em pre-authorizes} processing of personal data for around 600 vendors, even if those vendors are not present on the website, thus making consent being collected {\em for future and unforeseen potential processing}.
Therefore, such practice may violate the principles of transparency, fairness and minimization principles.
We hope our analysis of the IAB Europe TCF and the capability of CMPs to pre-register vendors that do not yet process personal data, will help policy makers to provide further guidance on the validity of such practice.
\\

\noindent
{\bf 
 Further recommendations are needed due to the decision-making power of \consentbanner\ providers.} %
In this article, we have analysed two most popular CMPs in the EU %
-- QuantCast and OneTrust-- and 
detected several scenarios when \consentbanner\ providers can be considered data controllers due to the enormous power of CMPs that can inject any type of additional functionality at any time in the banner, without the publisher being in position to technically know or oppose to it.
We hope that policy makers take these scenarios into account and provide recommendations for such providers (either withing or outside of IAB Europe TCF) identifying which practices render them as data controllers and in which conditions they will be recognized as data processors.

\section{Related work}
\label{sec:related}

Previous work analysing the role of CMPs in the advertising ecosystem have examined its technical functioning and interaction designs related to the applicable regulation, but have not inquired how they relate to their role as processors or controllers under the GDPR.

Degeling et al.~\cite{Dege-etal-19-NDSS} monitored the prevalence of CMPs on websites from January 2018 until May, when the GDPR came into effect, and measured an overall increase from
50.3\% to 69.9\% across all 28 EU Member States.
Taking a longer view, Hils et al. \cite{Hils-etal-20-IMC} showed how the rate of adoption doubled year over year between June 2018 and 2020, and that CMPs are mostly used by moderately popular websites (albeit with a long tail of small publishers).
Nouwens et al. \cite{Nouwens-etal-20-CHI} studied the use of dark patterns in the five most popular CMPs in the UK and estimated that only 11.8\% of banners meet minimum legal requirements for a valid consent (reject as easy as accept, no pre-checked boxes, and no implied consent). 

Focusing on the programmatic signals rather than user behaviour, Matte et al.~\cite{Matt-etal-20-IEEESP} analysed ~28,000 EU websites and found that 141 websites register positive consent even if the user has not made their choice and 27 websites store a positive consent even if the user has explicitly opted out.
Additionally, Matte et al.~\cite{Matt-etal-20-APF}
discuss the purposes and legal basis pre-defined by the IAB Europe and suggest that several purposes might not be specific or explicit enough to guarantee a valid legal basis, and that a large portion of purposes should require consent but are allowed by the TCF to be gathered on the basis of legitimate interest. 

Data protection authorities across EU Member States have also reacted to the role and responsibility of CMPs, and issued various guidances.
The Spanish DPA~\cite{AEPD-Guid-21} asserts that as long as CMPs comply with the requirements for consent, they shall be deemed an appropriate tool. It recommends that CMPs ``\textit{ must be submitted to audits or other inspections in order to verify that (...) requirements are complied with}''.
The Irish DPA~\cite{Irish-Guid-20} reiterates CMPs should be careful to avoid non-compliant designs already explicated as part of GDPR texts (e.g., pre-ticked boxes) and emphasises their  accountability and transparency obligations (i.e., consent records) %
The Danish DPA asserts that whenever any entity integrates content from any third party (including CMPs), it is particularly important to be aware of its role in relation to its processing of personal data that takes place~\cite{DanishDPA-2019}.

\fi

\section{ Conclusion}
\label{sec:conclusions}
In this paper we discussed the requirements for CMPs to be qualified as processors and as controllers and concluded that such status has to be assessed with regard to each specific data processing activity. 
From an empirical analysis we concluded that CMPs assume the role of controllers, and thus should be responsible for their processing activities, in four scenarios: i) when including additional processing activities in their tool, ii) when they perform scanning and pre-sorting of tracking technologies, iii) when they include
third-party vendors by default, and finally iv) when they deploy interface manipulative design strategies.

\section*{Acknowledgements}

We would like to thank Daniel Woods, Triin Siil, Johnny Ryan and anonymous reviewers of ConPro'21 and APF'21 for useful comments and feedback that has lead to this paper.
This work has been partially supported by the ANR JCJC project PrivaWeb (ANR-18-CE39-0008) and by the Inria DATA4US Exploratory Action project.

\bibliographystyle{splncs04}
\bibliography{bib/articles,bib/urls,bib/articles-M,bib/urls-M}

\iffalse
%
%
%
\onecolumn
\begin{appendices}

\subsection{Additional figures for Section~\ref{sec:controllers}} 
\label{appendix:cookiepro-banner}

\begin{figure}[hb!]
  \centering
    \includegraphics[width=0.95\textwidth]{fig/cookiepro-banner}
    \caption{2\textsuperscript{nd} layer of the default consent pop-up provided by CookiePro Free IAB TCF 2.0 CMP Builder (owned by OneTrust). [Captured on 13 Jan. 2021]. On the left, the top level of the page, displaying the ``Accept All'' button. On the right, the bottom of the same screen, displaying the ``Reject All'' and ``Confirm My Choices'' buttons, so the user needs to scroll down in order to see them.}
  \label{fig:cookiepro-banner}
\end{figure}

\begin{figure}[!hpb]
    \centering
    \includegraphics[width=0.95\textwidth]{fig/cookiepro_2.png}
    \caption{CookiePro's configuration back-end designed for the publisher, when logged. After completing a scan for trackers on the publisher's website, this screen shows the trackers that were found
    together with a category they are assigned with.}
    %
    \label{fig:cookiepro_scanresults}
\end{figure}

\iffull
\begin{figure}[htb]
  \centering
    \includegraphics[width=0.95\textwidth]{fig/quantcast-vendors-2.png}
    \caption{Installation process of QuantCast CMP [Captured on 5 Feb. 2021].  A publisher has to manually search and { exclude one-by-one the vendors from the list of 671 vendors registered in  IAB Europe TCF.}}
  \label{fig:vendors}
\end{figure}
\fi 

\end{appendices}

\fi

\end{document}